\documentclass[aps,prl,preprint,groupedaddress,nofootinbib]{revtex4-1}
\usepackage{graphicx}
\begin{document}


\title{\large  
A Unified Model for Inflation, pseudo-Goldstone Dark Matter, Neutrino Mass and Baryogenesis
}


\author{\bf Rabindra N. Mohapatra$^a$}
\author{Nobuchika Okada$^b$}
\affiliation{}
\affiliation{$^a$ Maryland Center for Fundamental Physics and Department of Physics, University of Maryland, College Park, Maryland 20742, USA}
\affiliation{$^b$ Department of Physics, University of Alabama, Tuscaloosa, Alabama 35487, USA}


\date{\today}

\begin{abstract} 
We present a unified theory of inflation, neutrino mass, baryogenesis and dark matter where  global lepton number symmetry and its breaking play a crucial  role. 
The basic idea is to use a lepton number carrying complex scalar field as the inflaton as well as the field that implements Affleck-Dine (AD)  leptogenesis. 
Dark matter is the massive majoron which is a pseudo-Goldstone boson, resulting from the  spontaneous breaking of lepton number symmetry 
  supplemented by explicit lepton number violation needed to implement AD leptogenesis.  
The  magnitude of the resulting $n_B/s$ in the model is related to the mass of the pseudo-Goldstone dark matter, 
  connecting two apparently disconnected cosmological observations.  
Inverse seesaw mechanism with lepton number breaking at low scale is crucial to prevent washout of the lepton asymmetry
  during the universe's evolution.  The model seems to provide an economical solution to several puzzles of the standard model of particle physics and cosmology in one stroke.

\end{abstract}

\maketitle

\section{1. Introduction} Some of the problems of particle physics and cosmology currently under great deal of scrutiny are: 
   (i) origin of neutrino masses, (ii) origin of matter in the universe, (iii) nature of dark matter and (iv) finally the origin of  the inflationary expansion 
   of the early universe. 
All these call for new ideas and scenarios of physics beyond the standard model and will broaden the frontier of our knowledge 
   regarding the matter and forces as well as the evolution of the universe.
The current landscape of beyond the standard model (BSM) physics  includes many proposals that provide solutions to one or more of these problems. 
Our goal in this paper is to provide a simple unified extension of the standard model (SM) of particle physics that provides resolutions of all these problems in an interconnected manner. The framework is based on the Affleck-Dine (AD)  proposal for baryogenesis ~\cite{AD} where the inflaton field and the AD field are one and the same, thus  providing first unification of two different phenomena into one~\cite{Cline:2019fxx, Charng:2008ke, Hertzberg:2013jba, Takeda:2014eoa, Lin:2020lmr, Stubbs, russian, Kawasaki:2020xyf, Barrie:2021mwi, nobu}.  
We endow the same AD and inflaton field with a global $B-L$ quantum number which is broken to generate neutrino masses via the inverse seesaw mechanism~\cite{ISS1, ISS2}.  
We find the inverse seesaw to be the chosen path to make our scenario consistent with the generation of baryon asymmetry.
The associated singlet majoron field \cite{majoron}, which acquires mass due to the explicit breaking inherent to the AD leptogenesis mechanism,
  plays the role of a pseudo-Goldstone dark matter recently discussed in the literature 
   \cite{pgDM0,pgDM1,pgDM2,pgDM3,pgDM4,pgDM5,pgDM6,pgDM7,pgDM8,pgDM9}.   
The mass of the DM ($m_{DM}$) is connected to the magnitude of baryon asymmetry $n_B/s$.
Thus in some sense the AD field plays the central unifying role behind our proposal for solving inflation, dark matter, baryogenesis and neutrino mass problems of the SM. 

The dark matter in our model is an unstable  particle with its lifetime above $10^{26}$ seconds consistent with all other constraints on the model. Its high degree of stability is guaranteed by the approximate $Z_2$ symmetry  in the model which is related to the $B-L$ symmetry (just like in the case of SUSY models~\cite{RNM} though in a different way).   
The pseudo-Goldstone nature of the dark matter also explains the lack of signal in underground search experiments for dark matter, 
  even though its mass  is in multi-GeV range. This last point has already been discussed in the literature.

To avoid washout of the lepton asymmetry resulting from the AD mechanism, it is necessary to generate small neutrino masses via the inverse seesaw mechanism~\cite{ISS1,ISS2} as noted. The essential point is that inverse seesaw corresponds to a low scale for lepton number breaking, an inherent property of the mechanism and in our model, this low scale makes it possible to maintain lepton number conservation till very low ($\sim$ 10 GeV) temperature helping to avoid the washout of lepton asymmetry generated before. 

This paper is organized as follows: in sec.~2, we present an outline of the model and isolate its symmetries; 
in sec~3, we discuss the scalar spectrum of the model necessary to understand the origin of the  pseudo-Goldstone dark matter;  
in sec.~4, we discuss the evolution of the universe in this picture and leptogenesis. 
In sec.~5, we focus on further details of the dark matter such as its lifetime and relic density generation. 
Sec.~6 is devoted to a summary of the paper.

\section{2. The model} 
\begin{table}[t]
\begin{center}
\begin{tabular}{|c||c||c|}
\hline
field & $U(1)_L$ & $U(1)_X$\\ \hline
$L$ & $+1$ & $0$\\
$N$ & $-1$ & $0$\\
$S$ & $0$ & $-1$\\\hline
$\Phi$ & $+1$ & $+1$\\
$\chi$ &  $-1/2$ & $-1/2$ \\
$\sigma$ & $0$ & $+2$\\
\hline
\end{tabular}
\end{center}
\caption{
Particle contents. 
The model has an extra $U(1)_X$ global symmetry in addition to 
the global Lepton number symmetry of $U(1)_L$. 
The SM lepton doublet is denoted as $L$. 
The generation index is suppressed. 
}
\label{tab1}
\end{table}

The model consists of three right handed neutrinos (RHNs) $N_{1,2,3}$ 
   and three more SM singlet neutral leptons $S_{1,2,3}$ are added to the SM
   along with a complex SM singlet scalar field $\Phi$ with $L=-1$, which will play the role of the inflaton/AD field,  
   another SM singlet scalar field $\chi$ with $L=+1/2$ to induce a vacuum expectation value (vev) 
   for the $\Phi$ field and finally a scalar field $\sigma$ whose vev will give a small Majorana mass to the singlet fermion $S$
   so that inverse seesaw can be implemented.  
The particle contents relevant to our discussion are listed in Table~\ref{tab1}.       
A linear combination of the imaginary parts of $\chi$ and $\Phi$ will play the role of dark matter. 
It will be a pseudo-Goldstone dark matter (denoted by pGDM).

The Lagrangian of the model is given by
\begin{eqnarray}
{\cal L} &=& {\cal L}_{SM}+{\cal L}_{inf} 
+ Y_D LHN+ Y_N \Phi NS + Y_\sigma \sigma SS \nonumber\\
&-& \left( m^2_\Phi |\Phi|^2+\lambda |\Phi|^4+\epsilon m^2_\Phi (\Phi^2+\Phi^{\dagger 2} ) \right) 
+\sqrt{2} \Lambda \left( \Phi \chi\chi + (\Phi \chi\chi)^\dagger \right) \nonumber\\
&-& 
\left( -m^2_{\chi} |\chi|^2 +\lambda_\chi |\chi|^4 \right) -\lambda_{mix} (\chi^\dagger\chi)(H^\dagger H) \nonumber\\
&-&
\left( -\mu^2_\sigma |\sigma|^2 +\lambda_\sigma |\sigma|^4 +\tilde{m}^2 (\sigma^2+ \sigma^{\dagger 2}) \right),
 \label{eq:L}
\end{eqnarray}
where  
$ {\cal L}_{SM}$ is the SM Lagrangian, 
${\cal L}_{inf}$ denotes the non-minimal $\Phi$ coupling to gravity that drives inflation~\cite{inf,inf1},
and $H$ is the Higgs doublet of the SM. 
As shown in Table~\ref{tab1}, the model has an extra global symmetry $U(1)_X$ in addition to $U(1)_L$,  
 which are explicitly broken by the $\epsilon m_\Phi^2$ and $\tilde{m}^2$ terms in the scalar potential.

 The first point to emphasize is that the model also has an extra broken discrete symmetry  in the scalar sector given by 
\begin{eqnarray}
\Phi\leftrightarrow \Phi^\dagger  \; \; {\rm and} \; \;\chi\leftrightarrow \chi^\dagger.
\end{eqnarray} 
Writing $\Phi=\frac{1}{\sqrt{2}}(\phi_1+i\phi_2)$ and $\chi=\frac{1}{\sqrt{2}}(\chi_1+i\chi_2)$, 
  we find that in the limit of $\epsilon =0$, there is  a remnant  $Z_2$ symmetry which transforms  
$(\phi_1, \chi_1)\to (\phi_1, \chi_1)$ and $(\phi_2, \chi_2) \to (-\phi_2, -\chi_2)$. 
This is analogous to R-parity in the MSSM and in our case, this symmetry keeps the linear combination
   of $\phi_2$ and $\chi_2$ fields highly stable as we see below. 
We note that this $Z_2$ symmetry is broken by the coupling of $\Phi$ to RHNs, which provides for the instability of $\phi_2$ and $\chi_2$. 
To proceed further, we discuss the vacuum state as well as the resulting scalar spectrum below.

Furthermore, below the temperature $T\leq  \langle \chi \rangle$, 
  the symmetry of the model reduces to $U(1)_{L-X}$ till the temperature when the $\sigma$ field picks up vev 
  and breaks all the global symmetries. 
The $\epsilon m_\Phi^2$ and $\tilde{m}^2$ terms explicitly break all the symmetries. 
As a result, the model has no domain wall problem.

\section{3. Scalar spectrum and Pseudo-Goldstone dark matter (pGDM)}
In order to analyze the scalar spectrum, we first display the vacuum state of the theory.  
Note that first $\chi_1$ acquires a vev due to the negative mass squared term for it. 
This then induces a vev for $\phi_1$ via the $\Phi\chi\chi$ term in the potential. We find by minimizing the potential that
\begin{eqnarray}
 && \langle \chi_1 \rangle  \equiv {v_\chi} =
   \frac{m_\chi m_\Phi \sqrt{1-2\epsilon}}{\sqrt{m^2_\Phi \lambda_\chi (1-2\epsilon)-2\Lambda^2} }, \nonumber\\
 && \langle\phi_1 \rangle  \equiv {v_\Phi} =\frac{\Lambda v^2_\chi}{m^2_\Phi (1-2\epsilon)}, \nonumber\\
 && \langle \phi_2 \rangle = \langle \chi_2 \rangle=0,
\end{eqnarray}
where we have assumed that $\lambda_{mix} \ll 1$. 
As just noted, the $\chi$ vev breaks a linear combination $L+X$  part of the two $U(1)$ symmetries 
  and leaves $L-X$ intact until the $\sigma$ field  acquires a vev. We choose $\langle \sigma \rangle$  below 100 GeV.
The effect of this is that as long as $\langle \sigma \rangle =0$, 
  there is an effective lepton number symmetry in the theory given by $U(1)_{L-X}$. 
For $T \geq \langle \sigma \rangle $, therefore, 
  the effective lepton number is conserved for the processes involving $N$ and $\bar{N}$.  
This helps to maintain any lepton asymmetry generated in earlier epochs of the universe from $\Phi$ decay. 
The vev of $\sigma$ breaks the final $L-X$ $U(1)$ symmetries and leaves a pseudo-Goldstone field 
  which picks up mass due to the $\tilde{m}^2$ term.  
In order to make the analysis simple, we assume that the $\sigma$ field does not mix with $\chi$ and $\Phi$ fields. 
The $\chi$ and $\Phi $ field components, however, mix with each other, the analysis of which is given below.

The mass matrices for the scalar fields need to studied to isolate the pseudo-Goldstone mode, 
  which will become the dark matter in our model, as stated above.  
We find the mass matrix for $(\chi_1, \phi_1)$ to be
\begin{eqnarray}
M^2_{R}~=~\left(\begin{array}{cc}2\lambda_\chi v^2_\chi &-2v_\chi \Lambda\\-2v_\chi\Lambda & m^2_\Phi (1-2\epsilon)\end{array}\right), 
\end{eqnarray}
and for $(\chi_2, \phi_2)$
\begin{eqnarray}
M^2_I~=~\left(\begin{array}{cc} 4\Lambda v_\phi & 2v_\chi \Lambda\\2v_\chi\Lambda & m^2_\Phi(1-2\epsilon)\end{array}\right).
\end{eqnarray}
The real part mass matrix has two positive eigenvalues if $\lambda_\chi m^2_\Phi (1-2\epsilon) -2\Lambda^2 > 0$. 
The determinant of the $M^2_I $ is given by 
\begin{eqnarray}
{\rm det}\left[ M^2_I \right] ~=~4\Lambda^2 v^2_\chi \left(\frac{1+2\epsilon}{1-2\epsilon}-1\right).
\end{eqnarray}
Note that as we set $\epsilon=0$, the determinant of $M^2_I$ vanishes and there is a massless boson which is the majoron. 
Since $\epsilon$ breaks $B-L$ symmetry explicitly, the lighter eigenvalue mass (majoron mass) denoted by $m_{DM}$ becomes
\begin{eqnarray}
m^2_{DM}~=~\frac{16\Lambda^2 v^2_\chi\epsilon}{4\Lambda v_\Phi +m^2_\Phi (1+2\epsilon)}.
\end{eqnarray}
The eigenstate corresponding to the DM is given by $\chi_{DM}=\cos\theta \chi_2+ \sin\theta \phi_2$ 
 with ${\rm tan} 2\theta = \frac{2v_\chi \Lambda}{m^2_\Phi (1+2\epsilon)-4\Lambda v_\Phi}\simeq \frac{2 \Lambda v_\chi}{m^2_\Phi}$
 in our benchmark set of parameters that will be given in the next section. 
 
Turning to the lepton sector, the RHN masses are given by $M_{N} = Y_N v_\Phi/\sqrt{2}$
  and the light neutrino masses are given by the inverse seesaw formula 
  $m_\nu\simeq M^T_D M_N^{-1}\mu M_N^{-1} M_D$, 
  where $M_D=Y_D v_{EW}/\sqrt{2}$ with the SM Higgs vev of $v_{EW}=246$ GeV,
  and $\mu= Y_\sigma \langle \sigma \rangle$ is a Majorana mass.

\section{4. Implications of the model} 
\begin{table}[t]
\begin{center}
\begin{tabular}{|c||c|}\hline
parameter &value\\\hline
$m_\Phi$& $10^6$ GeV\\
$v_\chi$ & $10^{15}$ GeV\\
$v_\Phi$ & $10^{11.5}$ GeV\\
$\Lambda$ & $10^{-7}$ \\
$\epsilon$ & $ 10^{-5}$ \\
$\theta$ & $10^{-4}$\\
$Y_{N}$ & $10^{-6.5}$\\
$M_{N}=Y_{N}v_\Phi/\sqrt{2}$ & $10^{5}$ GeV\\\hline
\end{tabular}
\end{center}
\caption{
A benchmark set of parameters that satisfies all the constraints considered in this section. 
}
\label{tab2}
\end{table}

In this section, after a brief review of the evolution of the universe in the model, 
  we discuss the origin of AD leptogenesis and associated issues.  
We illustrate that our model works with a benchmark set of parameters shown in Table~\ref{tab2}, 
  although there is a broader range of parameters where the model is viable.

\subsection{4a. Inflation and evolution of the AD field leading to lepton asymmetry} 
First we review the various stages in the evolution of the inflaton/AD field $\Phi$. 
We note that as in Ref.~\cite{nobu}, we adopt a non-minimal coupling of the $\Phi$ field to gravity to implement inflation. We do not repeat the detailed discussion of this which are given in Ref.~\cite{nobu} and we refer to this paper for the details of the various stages in the evolution of the universe as well as  the origin of $n_B/s$. 
The inflation is characterized by a parameter $\xi$  (which denotes coupling of $\Phi$ to the Ricci scalar) so that for $\Phi \geq M_P/\sqrt{\xi}$, the universe undergoes an inflationary phase. 
It fits all the Planck 2018 data on spectral index and the tensor-to-scalar ratio for $\xi\sim 1600$ for small $\lambda\sim 0.001$. 
The $\Phi$ slowly rolls down the potential and inflation comes to an end as $\Phi$ becomes less than $M_P/\sqrt{\xi}$. The $\Phi $ field then decreases like $1/a$, where $a$ is the scale factor of the universe, until its value is below $m_\Phi/\sqrt{\lambda_\Phi}$. Then the oscillation of the $\Phi$ field starts separately for its real and imaginary parts and they evolve starting from two random values for the two parts. 
This difference between the initial values of $\phi_1$ and $\phi_2$, introduces the CP violation required by the Sakharov's criterion
  for baryo/leptogenesis. 
The oscillation of the AD field leads to an asymmetry in the abundance of $N S$ and $\bar{N} \bar{S}$ 
  which is generated when the AD field decays as $\Phi \to N S$. 
Once $N$s are created, the universe is immediately thermalized with the plasma of SM particles 
  through the Yukawa interaction of $Y_D LHN$. 
We estimate the reheat temperature by $T_R\simeq \sqrt{\Gamma_{\Phi\to NS}M_P}$, 
  where $\Gamma_{\Phi\to NS}$ is the decay width of the inflaton/AD field, 
  and $M_P=2.4 \times 10^{18}$ GeV is the reduced Planck mass. 

\subsection{4b. AD leptogenesis,  inverse seesaw  and washout constraints}
As noted above, the difference between the initial values of $\phi_1$ and $\phi_2$ introduces the CP violation
  required by the Sakharov's criterion for baryo/leptogenesis and leads to lepton asymmetry
  when $\Phi$ decays to the Dirac RHNs via $\Phi\to NS$ process at the reheat temperature $T_R$ noted above. 
We choose parameters such that $T_R < 0.1 \, m_\Phi$.  
For our choice of $m_\Phi=10^6$ GeV in Table~2, it implies that $Y_N \simeq 10^{-6.5}$ for which $T_R\simeq 10^5$ GeV.  
First point to re-emphasize is that since lepton number breaking in inverse seesaw case occurs 
  below $\langle \sigma \rangle \simeq100$ GeV, the lepton number is conserved in all the processes involving $N$s 
  and the $NS$ asymmetry created by the inflaton/AD field decay gets transferred to the lepton asymmetry in the SM sector.  
However, one has to discuss the washout processes by explicit lepton number violating terms in the scalar potential
  and show that this asymmetry survives.

There are two sources of possible washout in our model: one for the $NS$ asymmetry due to the $\epsilon m_\Phi^2 \Phi^2$ term
  in the scalar potential that breaks lepton number by two units, and the second one that can wash out the SM lepton asymmetry 
  is the $\tilde{m}^2\sigma^2$ term since the $\sigma$ field connects to two $S$ fermions and leads to $X-L=4$ processes. 
These interactions must be out of equilibrium at and below the reheat temperature $T_R$ for the $NS$ asymmetry to survive
  and lepton asymmetry that lead to baryon asymmetry through the sphaleron transitions.  
To guarantee that  the $\epsilon m^2_\Phi \Phi^2$ term in the potential which mediates the dangerous process 
   $NS \leftrightarrow \bar{N}\bar{S}$  stays out of equilibrium at $T_R$,  we must satisfy
\begin{eqnarray}
  T_R^3 \times \frac{Y_N^4}{4 \pi} \frac{\epsilon^2 T_R^2}{m_\Phi^4} < H \simeq \sqrt{\frac{\pi^2}{90} g_{*} } \frac{T_R^2}{M_P},
\end{eqnarray}
where $g_* \simeq 100$ is the effective degrees of freedom of the SM thermal plasma. 
For our choice of $M_\Phi=10^6$ GeV, $T_R \simeq 10^5$ and $\epsilon= 10^{-5}$ in Table~\ref{tab2}, 
   we can see that this no-washout condition is satisfied. 

The second washout condition arises from the $\sigma$ interaction with the $\tilde{m}^2 \sigma^2$ term, 
   which mediates a dangerous process $SS \leftrightarrow \bar{S} \bar{S}$.
Assuming that the $\sigma$ mass is in the 100 GeV range, 
  we consider the washout condition on the parameters of the model 
  in two temperature regions: the first one is for $T > M_N$ and  
  the second one is for $T \leq M_N$.
The first condition is: 
\begin{eqnarray}
 T^3 \times \frac{Y_\sigma^4}{4 \pi} \frac{{\tilde m}^4}{T^6} < H \simeq \sqrt{\frac{\pi^2}{90} g_{*} } \frac{T^2}{M_P}. 
\end{eqnarray} 
This leads to 
\begin{eqnarray}
T  >   (Y_\sigma \tilde{m })\left(\frac{ M_P}{Y_\sigma \tilde{m}}\right)^{1/5}
\end{eqnarray}
for the process to be out of equilibrium. 
If we set $Y_\sigma =0.3$ and ${\tilde m}=30$ GeV, for example, we obtain $T > 10^4$ GeV, 
   so that this out-of-equilibrium condition is satisfied for a temperature $T > M_N=10^5$ GeV.
For the second case, $N$ is non-relativistic and the out-of-equilibrium condition is given by
\begin{eqnarray}
  (T M_N)^{3/2} e^{-\frac{M_N}{T}} \times \frac{Y_\sigma^4}{4 \pi} \frac{{\tilde m}^4}{M_N^6}  
  <  H \simeq \sqrt{\frac{\pi^2}{90} g_{*} } \frac{T^2}{M_P}. 
\end{eqnarray}
Since the number density of $N$ is exponentially suppressed for $T< M_N$, 
  the out-of-equilibrium condition is easily satisfied. 

We further note that in such a leptogenesis scenario, the lepton number to entropy ratio is given by~\cite{Stubbs} 
\begin{eqnarray}
\frac{n_B}{s}~\simeq \frac{T_R^3}{m^2_\Phi\epsilon M_P}.
\label{nB}
\end{eqnarray}
The set of benchmark in Table~\ref{tab2} does reproduce $n_B/s\simeq 10^{-10}$ as desired.

The final thing we have to discuss is the decay of the pseudo-Goldstone boson $\sigma$ in the early universe.  
For this, we include a coupling of $\sigma$ to the SM Higgs as follows: $\Lambda_\sigma \sigma H^\dagger H$. 
This generates a mixing between both the real and the imaginary parts of the $\sigma$ field to SM fields, 
  which clearly leads to effective $\sigma f \bar{f}$ couplings. 
This effective Yukawa couplings are large enough, so that the $ \sigma$ field does not survive below temperature $T$ 
  equal to its mass in the multi GeV range and does not affect the Big Bang Nucleosynthesis of the standard Big Bang cosmology.



\section{5. Pseudo-Goldstone Dark matter } 
We observed that the linear combination of $\chi_2$ and $\phi_2$ is a highly stable scalar field due to the weakly broken $Z_2$ symmetry of the model and can therefore play the role of dark matter. 
However in order to qualify as a viable dark matter, it must have a lifetime longer than $10^{26}$ seconds. 
Secondly it must have the right relic density. 
In this section we elaborate on both these points and show the viability of our scenario for the benchmark set of values of parameters in Table II. 
Notice that given the parameter choice, we find the dark matter mass to be $\sim $1 GeV using the formula given above  i.e.
\begin{eqnarray}
m_{DM}\simeq \frac{4\Lambda v_\chi \sqrt{\epsilon}}{m_\Phi}.
\end{eqnarray}
Note that the mass of the dark matter is connected to the amount of lepton asymmetry (see Eq.\ref{nB}) via the parameter $\epsilon$, thus connecting two apparently unrelated cosmological parameters.

\subsection{5a.  Dark matter lifetime} 
Let us discuss the lifetime of the dark matter. 
We denote the pGDM as $\chi_{DM}$ which is approximately identical to $\chi_2$.  
Its main decay mode is $\chi_{DM}\to \nu\nu$. 
The effective $\chi_{DM}\to \nu\nu$ coupling can be estimated as follows: 
The pGDM has a mixing with $\phi_2$ (denoted by $\theta$  above) through which it effectively couples with $N S$. 
The way that the effective $\chi_{DM}$ coupling to neutrinos arises is a bit subtle in the inverse seesaw case.
Note that in the limit of $\mu=0$,  the eigenstates of the $(\nu, N, S)$ mass matrix 
  are the state of $\nu \sin \psi +S \cos \psi$ pairing with $N$ to form a Dirac fermion 
  with mass $\sqrt{m^2_D+M^2_N}$ and a massless chiral fermion which is the physical neutrino mixed with $S$.  
Here, the mixing angle is given by $\psi \simeq m_D/M_N \ll$1.  
In this limit, we see that $S$ in the $NS$ final state contains the admixture massless neutrino 
   with a mixing angle $\psi$ while $N$ has no neutrino component. 
Hence, no $\chi_{DM}\to \nu\nu$ coupling arises in this limit. 
Once the $\mu$ term in the inverse seesaw matrix is generated, 
  the effective $\chi_{DM}\to \nu\nu$ Yukawa coupling ($Y_{DM})$ arises as
\begin{eqnarray}
Y_{DM} \simeq \theta \, \left( \frac{M_N}{v_\Phi} \right) \, \psi \, 
  \left(\frac{\mu m_D}{M^2_N} \right)  \simeq  \theta \, \frac{m_\nu}{v_{\Phi}} \simeq  \frac{m_\nu}{v_\chi}, 
\end{eqnarray}
 where we have used  $\theta\simeq \frac{\Lambda v_{\chi}}{m^2_\Phi} \simeq \frac{v_\Phi}{v_\chi}$, 
    and the inverse seesaw formula for the light neutrino mass $m_\nu$. 
%
We then estimate the pGDM lifetime as 
\begin{eqnarray}
\tau^{-1}_{DM} \simeq \frac{1}{4\pi} \left(\frac{ m_\nu}{v_\chi}\right)^2 \, m_{DM} \sim 10^{-27}~{\rm sec}^{-1}
\end{eqnarray}
for $m_\nu \sim 0.1$ eV and $m_{DM} \sim 1$ GeV, 
which is consistent with the lifetime constraint on a decaying dark matter from various astrophysical observations. 
For a recent reference, see Ref.~\cite{sinha}.  

\subsection{4a. Relic density of pGDM}
Let us now calculate the relic density of the pGDM of our model. 
We first note that the pGDM field in the polar basis is not directly coupled to SM fields.  
To see this we give below the polar basis Lagrangian for the $H$ and $\chi$ fields.
The relevant potential part can be written out from the Lagrangian of Eq.~(\ref{eq:L}) as follows:
\begin{eqnarray}
  V &\supset  &  
 -m^2_{\chi} |\chi|^2 +\lambda_\chi |\chi|^4  - \frac{1}{4} m^2\left( \chi^2 + (\chi^\dagger)^2 \right) 
 +\lambda_{mix} (\chi^\dagger\chi)(H^\dagger H) 
 \label{eq:potential}
\end{eqnarray}
where the parameter $m^2=\Lambda v_\Phi$ which results from substituting $\langle \Phi \rangle=v_\Phi/\sqrt{2}$
 in Eq.~(\ref{eq:L}) breaks the $U(1)_L$ symmetry softly.  
Going to the polar basis for parameterization of the $\chi$ field 
  i.e.~$\chi=\frac{1}{\sqrt{2}}(\rho+v_\chi)e^{i \frac{\varphi}{v_\chi}}$, 
  Eq.~(\ref{eq:potential}) is expressed as  
\begin{eqnarray}
  V &\supset  &  
-\frac{1}{2} m^2_{\chi} (\rho+v_\chi)^2 + \frac{1}{4} \lambda_\chi (\rho+v_\chi)^4  
 - \frac{1}{4} m^2 \left( \rho+v_\chi \right)^2 \left(1 - 2 \sin^2\left( \frac{\varphi}{v_\chi}\right)  \right)  \nonumber \\
&&  + \frac{1}{2} \lambda_{mix} \left( \rho+v_\chi \right)^2 (H^\dagger H) .
\end{eqnarray}
In the polar basis, the $\varphi$ field is identified with the pGDM field. 
Note that this pGDM is massless until ${m}^2$ is taken into account. 
Minimizing the above potential 
  we find the mass spectrum, $m^2_\rho=2\lambda_\chi v^2_\chi$ and $m^2_\varphi=m^2$. 
Here, we have assumed $\lambda_{mix} \ll 1$ and hence neglected the $\lambda_{mix}$ term 
  and the SM Higgs potential in the minimization. 
In the polar basis, the kinetic term of $\chi$ is expressed as 
\begin{eqnarray}
   (\partial_\mu \chi)^\dagger (\partial^\mu \chi) = \frac{1}{2}  (\partial_\mu \rho) (\partial^\mu \rho) 
   + \frac{1}{2} \left( 1+ \frac{\rho}{v_\chi} \right)^2 (\partial_\mu \varphi) (\partial^\mu \varphi) .
\end{eqnarray}
As mentioned above, the pGDM field $\varphi$ has no direct coupling to the SM Higgs field
 and it couples to it only via the $\rho$ field.

To estimate the relic density of the pGDM field, let us choose $\lambda_{mix} \ll 1$.
In this case, the absence of any interaction of the pGDM with the SM fields guarantees
  that it is not present in the thermal plasma of the SM particles in the early universe. 
Since $\lambda_{mix}$ is the coupling of the $\rho$ field with the SM Higgs field,    
  the $\rho$ field is also not in the thermal plasma either. 
Thus the pGDM is a feebly interacting dark matter and we need to make sure its relic density is generated
   via its interaction with the $\rho$ field. 
In our benchmark set of parameters, $v_\chi \gg T_R$ so that $m_\rho \gg T_R$ 
   unless $\lambda_\chi$ is extremely small.   
In this case, the dominant process for the pGDM production turns out to be the process 
  of $HH^\dagger \to \varphi\varphi$ via the mediation with the heavy $\rho$ field \cite{Abe:2020ldj}. 
This cross section is given by
\begin{eqnarray}
\sigma(HH^\dagger \to \varphi\varphi) \simeq \frac{\lambda^2_{mix}}{4 \pi}\frac{s}{m^4_\rho} 
\end{eqnarray}
for  $s \ll m_\rho^2$. 

The Boltzmann equation for the freeze-in \cite{freezein} pNGDM is given by
\begin{eqnarray}
\frac{dY}{dx}\simeq \frac{s(m_\varphi)}{H(m_\varphi)} \frac{\langle \sigma v \rangle}{x^2} Y^2_{eq},
\end{eqnarray}
where  $Y$ is the ratio of $\varphi$ number density $n$ to entropy density $s=\frac{2\pi^2}{45} g_* T^3$ 
  at temperature $T$,  $x=\frac{m_\varphi}{T}$, $s(m_\varphi)$ and $H(m_\varphi)$ are, respectively, 
  the entropy density and the Hubble parameter $H=\sqrt{\frac{\pi^2g_*}{90}}\frac{T^2}{M_P}$ at $T=m_\varphi$,  
  and the $Y_{eq}$ is $Y$ value when the pGDM is in termal equilibrium. 
Here, the thermal average of the DM production $\langle \sigma v \rangle$ is given by
\begin{eqnarray}
 \langle \sigma v \rangle \simeq (n_{eq})^{-2}\frac{T}{64\pi^4}\int_{4m^2_\varphi}^\infty ds \, 2 (s-4m^2_\varphi)
 \sigma(HH^\dagger \to \varphi\varphi) \sqrt{s} K_1  \left( \frac{\sqrt{s}}{T} \right),
\end{eqnarray}
where $K_1$ is the modified Bessel function of the 1st kind. 
Using $T\gg m_\varphi$, we get 
\begin{eqnarray}
  \langle \sigma v \rangle \simeq (n_{eq})^{-2}\frac{6}{\pi^5}\frac{\lambda^2_{mix}}{m^4_{\rho}} T^8. 
\end{eqnarray}
We can then calculate the relic density of dark matter by using the formula, 
\begin{eqnarray}
 \Omega_{DM}h^2 = \frac{m_\varphi Y(x \to \infty) s_0}{\rho_c/h^2} 
\end{eqnarray}
  with the entropy density $s_0=2890/{\rm cm}^3$ 
  and $\rho_c/h^2 = 1.05 \times 10^{-5} \, {\rm GeV}/{\rm cm}^3$ is the critical density at present.  
We reproduce the observed dark matter relic density of $\Omega_{DM}  h^2=0.12$ \cite{Planck:2018vyg}
  for the parameter choice below: 
\begin{eqnarray}
\lambda_{mix}\simeq 6.7 \times 10^{-8}\left(\frac{m_\rho}{10^6 \, {\rm GeV}}\right)^2 \left(\frac{1 \,{\rm GeV}}{m_\varphi}\right)^{1/2} \left(\frac{10^5 \, {\rm GeV}}{T_R}\right)^{3/2}.
\end{eqnarray}
These parameters are all in the range of the benchmark points in the Table~\ref{tab2}. 
Thus our model can explain the dark matter of the universe. 

\section{6. Summary} 
We have presented a simple  extension of the standard model that provides a unified explanation of several of its puzzles i.e. neutrino masses, dark matter compatible with current direct detection constraints, inflation and baryogegenesis via the Affleck-Dine mechanism. The model is quite economical in the sense that it adds only three right handed neutrinos, three other heavy singlet fermions which are the pseudo-Dirac partners of the RHN in the inverse seesaw explanation of the small neutrino masses, supplemented by  three lepton number carrying complex scalars bosons that play an important role in inflation, baryogenesis and dark matter physics. The model  parameters are highly constrained by the requirements of right physics. We also find it interesting that  the amount of baryon asymmetry in the model is intimately connected to the mass of the dark matter keeping it in the a GeV range. 
We demonstrate the viability of our model with a benchmark set of parameters shown in Table II. 
Clearly the model is viable in a domain of parameters around that.

\section*{Acknowledgement}
The work of R.N.M. is supported by the US National Science Foundation grant no.~PHY-1914631 and  the work of N.O. is supported by the US Department of Energy grant no.~DE-SC0012447.

\end{document}